\documentclass[hyper]{JHEP} 

\usepackage{epsfig}

\newcommand\fverb{\setbox\pippobox=\hbox\bgroup\verb}

\newcommand\fverbdo{\egroup\medskip\noindent%

            \fbox{\unhbox\pippobox}\ }

\newcommand\fverbit{\egroup\item[\fbox{\unhbox\pippobox}]}

\newbox\pippobox

\title{Note About  Unstable
M3-brane Action}
\author{J. Kluso\v{n}\\
Department of
Theoretical Physics and Astrophysics\\
Faculty of Science, Masaryk University\\
Kotl\'{a}\v{r}sk\'{a} 2, 611 37, Brno\\
Czech Republic\\

E-mail: \email{klu@physics.muni.cz}}

\preprint{arXiv:0810.0585}

 \abstract{We construct an action for
  unstable M3-brane in M-theory.
 We argue that in order to find
 M3-brane action that upon direct
 dimensional reduction leads to non-BPS
D3-brane action in type IIA theory we
have to presume that the background
possesses Killing isometry and that
this isometry has to be gauged on the
world-volume of M3-brane. Then we
construct singular tachyon kink on
world-volume of M3-brane and we show
that the dynamics of  resulting
topological defect is governed by an
action that upon direct dimensional
reduction leads to D2-brane in type IIA
theory and that  is equivalent to
standard M2-brane action in flat
background.}

 \keywords{M-theory}

\newcommand{\calg}{\mathcal{G}}
\newcommand{\calgi}{\left(\calg^{-1}\right)}

\newcommand{\hC}{\hat{C}}
\newcommand{\hg}{\hat{g}}
\newcommand{\hX}{\hat{X}}
\newcommand{\hsigma}{\hat{\sigma}}
\newcommand{\hrho}{\hat{\rho}}
\newcommand{\hx}{\hat{x}}

\newcommand{\hh}{\hat{h}}
\newcommand{\homega}{\hat{\omega}}
\newcommand{\mK}{\mathcal{K}}
\newcommand{\hmK}{\hat{\mK}}
\newcommand{\hA}{\hat{A}}
\newcommand{\mF}{\mathcal{F}}
\newcommand{\hmF}{\hat{\mF}}
\newcommand{\hD}{\hat{D}}
\newcommand{\hb}{\hat{b}}
\newcommand{\ha}{\hat{a}}
\newcommand{\bai}{\left(\ba^{-1}\right)}

\newcommand{\hk}{\hat{k}}

\def \bAi{\left(\mathbf{A}^{-1}\right)}

\def \bA{\mathbf{A}}

\newcommand{\ba}{\mathbf{a}}

\newcommand{\mL}{\mathcal{L}}

\begin{document}
\section{Introduction}\label{first}
Study of unstable  systems is an
important direction in string theory
research, for review and extensive list
of references, see
\cite{Sen:1999mg,Sen:2004nf}. In the
context of two superstring theories,
type IIA and type IIB, we have two
unstable systems: The first one is
system that consists Dp-brane and
anti-Dp-brane pair with $p$ is
odd(even) in type IIA(type IIB) theory.
The second one is a single unstable
Dp-brane where now $p$ is even(odd) in
type IIA(type IIB) theory.

These systems have rich and interesting
 dynamics. In particular, it is well known that the
 tachyon condensation on these objects leads
 to an emergence either stable Dp-branes or
 fundamental strings. Further,
 it is  remarkable fact that many these
phenomena can be described in the language
of tachyon effective action
\footnote{For some works considering effective
action description of an unstable systems, see
\cite{Garousi:2007fk,Garousi:2007fn,Cho:2007jf,Kluson:2007hb,Kluson:2006xi,
Kluson:2005fj,Kluson:2005hd,Garousi:2005zb,Kim:2005hea,Garousi:2004rd,
Banerjee:2004cw,Kluson:2004qy,Yee:2004ec,Niarchos:2004rw,Sen:2003zf,
Fotopoulos:2003yt,Smedback:2003ur,Kim:2003uc,Kwon:2003qn,Kim:2003ma,
Sen:2003bc,Brax:2003rs,Kim:2003ina,Okuyama:2003wm,Kutasov:2003er,
Garousi:2003pv,Yi:1999hd,Sen:2002vv,Sen:2002in,Sen:2002nu,Minahan:2000tf,Sen:2000kd,
Sen:2000vc,Mukhopadhyay:2002en,Gibbons:2000hf,Arutyunov:2000pe,Hashimoto:2001rk,Lambert:2001fa,
Kutasov:2000aq,Minahan:2002if,Lambert:2002hk,Alishahiha:2000du,Minahan:2000tg,Kluson:2000iy,
Bergshoeff:2000dq,Garousi:2002wq,Garousi:2000tr}.}.

On the other hand it is well known that
all stable D-branes, fundamental
strings and other extended objects in
string theories arise through
dimensional reductions from M2 and
M5-branes in M-theory \footnote{For
review and extensive list of earlier
papers, see \cite{Townsend:1996xj}.}.
Then it is natural to ask the question
whether it is possible to formulate
tachyon effective actions for M-theory
unstable objects that could be
interpreted as pre-images  of unstable
D-branes in super string theories.
An existence of such objects
 was predicted in very
interesting papers
\cite{Intriligator:2000pk} and
\cite{Houart:2000vm,Houart:1999rf}.
 We will not try to construct actions  for all
unstable systems discussed in this
paper. We rather restrict ourselves to
the case of an unstable M3-brane that
can be directly related to M2-brane. In
fact, it was argued in
\cite{Intriligator:2000pk}
 that the main property of
this M3-brane is that the  tachyon kink
solution  on the world-volume of
M3-brane should lead to an emergence of
M2-brane.
 Then motivated by success of
the effective action description of the
tachyon kink configuration
\cite{Sen:2003tm,Kluson:2005fj,Kluson:2005hd}
we try to formulate M3-brane effective
action that through the  tachyon kink
solution  leads to an action for stable
M2-brane \footnote{We are also
motivated by
  recent  progress in the study
of effective action for $N$ M2-branes
\cite{Bagger:2007vi,Bagger:2007jr}
since we would like to see whether it
is possible to find system of unstable
$N$ M3-branes that through the kink
tachyon condensation leads to the
effective action for $N$ M2-branes.}.

However we will see that there are some
subtleties with construction of an
action for such an object. More
concretely, we would like to find
M3-brane effective action that has two
following properties: 1. The singular
tachyon kink condensation leads to an
emergence of M2-brane action 2: The
dimensional reduction of M3-brane
action leads to an unstable D3-brane
action in type IIA theory.
Unfortunately it turns out that we are
not able to obey these two conditions
simultaneously. More precisely, in the
next section we formulate M3-brane
effective action with the property that
the singular tachyon kink condensation
on its world-volume leads to ordinary
M2-brane action. On the other hand,
when we  perform the dimensional
reduction to type IIA theory background
we derive the form of the action that
is different from the tachyon effective
action for unstable D3-brane that is
clearly unsatisfactory property of
given action.

Then in order to find M3-brane action
that reduces to non-BPS D3-brane action
in type IIA theory we  start our
analysis  in opposite direction.
However it turns out  that the
construction of given M3-brane action
demands an existence of isometry of
M-theory supergravity background that
has to be gauged on the world-volume of
M3-brane action.  Then it is clear that
singular tachyon condensation on the
world-volume of such a M3-brane leads
to gauged version of M2-brane action.
Then we study properties of this
M2-brane action and we find that the
direct dimensional reduction leads to
D2-brane action. Finally we show that
in flat background this action is
equivalent to standard M2-brane action.

The organization of this paper is as
follows. In  next section
(\ref{second}) we introduce the first
form of M3-brane action and show that
the singular tachyon kink leads to
M2-brane action. Then in section
(\ref{third}) we perform direct
dimensional reduction in this action.
In section (\ref{fourth}) we propose
alternative form of M3-brane action
that upon direct dimensional reduction
leads to an unstable D3-brane action.
We discuss its double dimensional
reduction in section (\ref{fifth}) and
argue that it leads to unstable
NS2-brane. Then in section
(\ref{sixth}) we also construct tachyon
kink solution on its world-volume and
we argue that the dynamics of the kink
is governed by gauged M2-brane action.
We study properties of this action in
section (\ref{seventh}). Finally in
conclusion (\ref{cons}) we outline our
results and suggest their possible
extension.
\section{Naive Form of  Unstable M3-brane
Action}\label{second} In this section
we propose the first form of an
unstable M3-brane action with the
property that the singular tachyon kink
on its world-volume corresponds
 to a stable M2-brane
action in general background. In fact,
with analogy with unstable D3-brane in
type IIA theory we propose the action
for this unstable M3-brane in the form
 \footnote{We use following
\cite{Bergshoeff:1998ef} conventions for hats. Hats
on target space fields
 indicate they are $11$-dimensional. Absence
of hats indicates they are
$10$-dimensional. Further, capital
$M,N,\dots$ label $11$-dimensional
indices, $m,n$ label $10$-dimensional
indices.}
\begin{equation}\label{M3act}
S=-\int d^4\xi V(T) \sqrt{-\det \bA}+
\int V(T)dT \wedge C \ ,
\end{equation}
where
\begin{eqnarray}
\bA_{\mu\nu}= \hg_{MN}
\partial_\mu \hX^M\partial_\nu
\hX^N+\partial_\mu T\partial_\nu T \ ,
\nonumber \\
\end{eqnarray}
and where $\xi^\mu \ , \mu,\nu=0,\dots,3$ label
world-volume coordinates of M3-brane. Further,
$V(T)$ is the tachyon potential that is
even function of $T$ with the property that
$V(T=\pm \infty)=0$ and $V(T=0)=\tau_{M3}$,
where $\tau_{M3}$ is M3-brane tension.

Let us now consider tachyon kink
solution on the world-volume of the
action (\ref{M3act}).
 To do this we
closely follow similar analysis
performed in case of unstable Dp-branes
in general background
\cite{Sen:2003tm,Kluson:2005fj,Kluson:2005hd}.
We start with the ansatz for
world-volume modes
\begin{eqnarray}\label{ansM3}
 T &=&f(a(\xi^3 - t(\xi^\alpha))) , \quad
 \nonumber \\
\hX^M&=& \hx^M(\xi^\alpha) \ ,
\nonumber \\
\end{eqnarray}
where $\xi^\alpha, \alpha,\beta=0,1,2$
are coordinates tangential to the
world-volume of the kink. The function
$f$ introduced in (\ref{ansM3})
 satisfies
\begin{equation}\label{fu}
  f(-u) =
-f(u) \ ,  \quad f'(u) > 0 \quad  \forall u \ ,
\quad f(±\infty) = ±\infty
\end{equation}
 but is otherwise
an arbitrary function of the argument
$u$. $a$ is a constant that we shall
take to $\infty$ in the end. In this
limit we have $T =\infty$  for  $\xi^3
> 0$  and $T = -\infty$  for $\xi^3 < 0$.

 The first goal of our
analysis is to shown that the action
(\ref{M3act}) evaluated on the ansatz
(\ref{ansM3}) leads to the standard
M2-brane action in general background
\begin{eqnarray}\label{M2act}
S&=&S_{M2}+S_{WZ} \ , \nonumber \\
S_{M2}&=& -T_{M2} \int d^3\xi
\sqrt{-\det\ba} \ ,\nonumber \\
 S_{WZ}&=&-\frac{1}{3!}T_{M2} \int
d^3\xi \epsilon^{\alpha\beta\gamma}
\hC_{\alpha\beta\gamma} \ , \nonumber
\\
 \ba_{\alpha\beta}&=&\partial_\alpha
\hx^M\partial_\beta \hx^N
\hg_{MN} \ , \quad  \hC_{\alpha\beta\gamma}=
\hC_{MNK}\partial_\alpha \hx^M\partial_\beta \hx^N
\partial_\gamma \hx^K \    \nonumber \\
 \end{eqnarray}
 that implies following
equations of motion for $\hx^M$
\begin{eqnarray}\label{eqx1}
-\frac{1}{2}\frac{\delta
\hg_{MN}}{\delta \hx^K}
\partial_\alpha \hx^M\partial_\beta \hx^N
\bai^{\beta\alpha}\sqrt{-\det\ba}+
\nonumber \\
+
\partial_\alpha\left[
\hg_{KM}\partial_\beta \hx^M
\bai^{\beta\alpha}_S
\sqrt{-\det\ba}\right]+\tilde{J}_K=0  \ ,
\nonumber \\
\end{eqnarray}
where $\tilde{J}_K=\frac{\delta}{\delta
\hx^K}
 S_{WZ}$ and where for letter purposes
we introduced
 the symmetric and
antisymmetric form of the matrix $\bai$
\begin{equation}
\bai_S^{\alpha\beta}
=\frac{1}{2}\left(\bai^{\alpha\beta}+\bai^{\beta\alpha}\right)
\ , \quad \bai_A^{\alpha\beta}=
\frac{1}{2}\left(\bai^{\alpha\beta}-\bai^{\beta\alpha}\right)
\ .
\end{equation}
As the first step we insert the ansatz
(\ref{ansM3}) to the matrix $\bA$ and we
obtain
\begin{eqnarray}
\bA_{\alpha\beta}&=& \partial_{\alpha}\hx^M
\partial_{\beta}\hx^N\hg_{MN}+
a^2f'^2\partial_\alpha t\partial_\beta
t \ , \nonumber \\
 \bA_{\alpha
3}&=&-a^2f'^2\partial_\alpha
t \ , \quad \bA_{3\beta}=-a^2f'^2\partial_\beta t  \ ,
\quad \bA_{33}=a^2f'^2 \
\nonumber \\
\end{eqnarray}
and consequently
\begin{equation}\label{detbA}
\det \bA= a^2f'^2 \det
\ba_{\alpha\beta}
\end{equation}
and
\begin{eqnarray}
\bAi^{33}&=&\frac{1}
{a^2f'^2}+\partial_\alpha
t\bai^{\alpha\beta}
\partial_\beta t \ , \quad
\bAi^{\alpha 3}= \bai^{\alpha\gamma}
\partial_\gamma t \ , \nonumber \\
\bAi^{3\beta}&=&
\partial_\gamma t \bai^{\gamma\beta}
 \ , \quad \quad
\bAi^{\alpha\beta}=\bai^{\alpha\beta} \ .
\nonumber \\
\end{eqnarray}
Then with the help of (\ref{detbA}) we
obtain that the action (\ref{M3act})
evaluated on the ansatz  (\ref{ansM3})
reduces into action (\ref{M2act}) in
the limit $a\rightarrow \infty$
\begin{eqnarray}
S_{M3}&=&- \int d\xi^3 d^3\xi
 a
f' V \sqrt{-\det \ba}
+\frac{1}{3!}\int d\xi^3 af'V d^3\xi
\epsilon^{\alpha\beta\gamma}\hC_{\alpha\beta\gamma}=
\nonumber \\
&=&
-T_{M2} \int
d^3\xi \sqrt{-\det\ba}+T_{M2}
\int \hC   \ , \nonumber \\
\end{eqnarray}
where we make standard presumption that
\begin{equation}
T_{M2}=\int dx af'(ax)V(f(ax))= \int du
V(u) \ .
\end{equation}
As the next step we show that the equations
of motion for $T$ and $\hX$ are obeyed
for the ansatz (\ref{ansM3}) on condition
that the modes $\hx^M$ obey the equations
of motion (\ref{eqx1}).
In fact, it is easy to determine
from (\ref{M3act}) the equation
of motion for $T$
\begin{equation}\label{eqT}
-V'(T)\sqrt{-\det\bA}+
\partial_\mu
\left[V(T)\partial_\nu T
\bAi^{\nu\mu}_S\sqrt{-\det\bA}\right]+J_T=0 \ ,
\end{equation}
where $J_T=\frac{\delta}{\delta T}
S_{WZ}$ is the source current derived
from varying the Wess-Zumino term.
Now using $\bA_{\mu\nu}\bAi^{\nu\rho}=
\delta_\mu^\rho$ we obtain
\begin{eqnarray}
\bA_{33}\bAi^{3\mu}+
\bA_{3\alpha}\bAi^{\alpha \mu}
=\delta^\mu_3 \Rightarrow  \nonumber \\
\bAi^{3\mu}_S-\partial_\alpha t\bAi^{\alpha\mu}_S
=\frac{1}{a^2f'^2}\delta^\mu_3 \nonumber \\
\end{eqnarray}
and consequently
\begin{eqnarray}
&-&V'(T)\sqrt{-\det\bA}+
\partial_\mu
\left[V(T)\partial_\nu T
\bAi^{\nu\mu}_S\sqrt{-\det\bA}\right]=\nonumber \\
&=&-V'(T)af'\sqrt{-\det\ba}+
\partial_\mu
\left[V(T)
a^2f'^2
(\bAi^{3\mu}_S-\partial_\alpha t\bAi^{\alpha \mu}_S
)\sqrt{-\det\ba}\right]
=0\nonumber \\
\end{eqnarray}
for any $t$. In fact, it can be also
easily shown that the tachyon current
$J_T$ is zero. In other words the
tachyon equation of motion is obeyed
for any $t$.

Let us now turn to the equations of
motion for the scalar modes $\hX^M$
that arise from the action (\ref{M3act})
\begin{eqnarray}\label{eqX}
&-&\frac{1}{2}
V\frac{\delta \hg_{MN}}{\delta
\hX^K}
\partial_\mu \hX^M\partial_\nu \hX^N
\bAi^{\nu\mu}\sqrt{-\det\bA}+
\nonumber \\
&+&
\partial_\mu\left[V
\hg_{KM}\partial_\nu \hX^M
\bAi^{\nu\mu}_S
\sqrt{-\det\bA}\right]+J_K=0  \ ,
\nonumber \\
\end{eqnarray}
where $J_K=\frac{\delta S_{WZ}}{\delta
\hX^K}$. Inserting the ansatz
(\ref{ansM3}) to the equations above we
obtain
\begin{eqnarray}\label{eqXi}
&-&\frac{1}{2}Vaf' \frac{\delta
\hg_{MN}}{\delta \hx^K}
\partial_\alpha \hx^M\partial_\beta
\hx^N\bai^{\alpha\beta}\sqrt{-\det\ba}
 +\nonumber \\
&+&Vaf'
\partial_\alpha \left[
\hg_{KM}\partial_\beta \hx^M
\bai^{\beta\alpha}_S
\sqrt{-\det\ba}\right]+J_K=\nonumber \\
&=&
 af'V[e.o.m.(\ref{eqx})]=0 \  \nonumber \\
\end{eqnarray}
using the fact that
\begin{eqnarray}
\partial_\alpha
[aVf']=-\partial_3[aVf']\partial_\alpha
t \  \nonumber \\
\end{eqnarray}
and also
\begin{eqnarray}
J_K=-V(T)\epsilon^{\mu_1\dots\mu_4}
\partial_{\mu_1}T\frac{\delta
\hC_{\mu_2\dots\mu_4}}{\delta\hx^K}=
-V(f)af'\epsilon^{\alpha\beta\gamma}
\frac{\delta
\hC_{\alpha\beta\gamma}}{\delta\hx^K}
=af'V\tilde{J}_K \ . \nonumber \\
\end{eqnarray}
Let us outline our result. We derived
 that the dynamics of the
tachyon kink is governed by the
equations of motion (\ref{eqx1}). In
other words we shown that the tachyon
kink on the world-volume of M3-brane
corresponds to stable M2-brane. This is
very nice result and in some sense
supports the proposal that
(\ref{M3act}) correctly describes
unstable M3-brane. On the other hand
as we show below
 dimensional reduction of this
M3-brane does not lead to the correct form
of an unstable D3-brane action.
\section{Dimensional Reduction}\label{third}
In this section we  test the properties
of M3-brane action (\ref{M3act})
further when we study the direct
dimensional reduction of given action.
 To do
this we will presume that  $D=11$
background has a $U(1)$ isometry with
Killing vector field $\hk^M$, such that
Lie derivative of all target space
fields with respect to $\hk$ vanish:
\begin{equation}
\mL_{\hk} \hg_{MN} =\mL_{\hk}
\hC_{MNK}=0 \ . \end{equation}
 In adapted coordinates $\hx^M =
(x^m, z)$ for which $\hk
=\frac{\partial}{\partial z} \ ,
\hk^M=\delta^M_z$ we can write the D=11
bosonic fields as
\begin{eqnarray}\label{ansdr}
\hg_{mn}&=&e^{-\frac{2}{3}\phi}g_{mn}
+e^{\frac{4}{3}\phi}C^{(1)}_m C^{(1)}_n
\ ,
\nonumber \\
\hg_{mz}&=&-e^{\frac{4}{3}\phi}C^{(1)}_m
\ , \quad  \hg_{zz}=e^{\frac{4}{3}\phi}
\ ,
\nonumber \\
\hC^{(3)}_{mnk}&=&C^{(3)}_{mnk} \ , \quad
(i_{\hk} \hC)_{mn}=B_{mn} \ ,
\nonumber \\
\end{eqnarray}
where we have $NS\otimes NS$ fields
$(\phi,g_{mn},B_{mn})$ and $R\otimes R$
fields $(C^{(1)}_m,C^{(3)}_{mnk})$. For this ansatz
we obtain
\begin{eqnarray}
\bA_{\mu\nu}&=&
e^{-\frac{2}{3}\phi} g_{mn}\partial_\mu
X^m\partial_\nu X^n+\partial_\mu
T\partial_\nu T+ \nonumber \\
&+& e^{\frac{4}{3}\phi} (\partial_\mu
X^m C_m^{(1)}-\partial_\mu Z)
(\partial_\nu X^n
C_n^{(1)}-\partial_\nu Z)=
\nonumber \\
&=&e^{-\frac{2}{3}\phi}
[g_{mn}\partial_\mu X^m\partial_\nu
X^n+ e^{\frac{2}{3}\phi}\partial_\mu T
\partial_\nu T+e^{2\phi}Y_\mu Y_\nu]
=\nonumber \\
&=&e^{-\frac{2}{3}\phi}
\calg_{\mu\rho}[ \delta^{\rho}_\nu+
e^{2\phi} \calgi^{\rho\sigma}Y_\sigma
Y_\nu] \ , \nonumber \\
\end{eqnarray}
where we defined
\begin{equation}\label{calgY}
\calg_{\mu\nu}=g_{mn}\partial_\mu X^m
\partial_\nu
X^n+e^{\frac{2}{3}\phi}\partial_\mu
T\partial_\nu T \ , \quad
Y_\mu=\partial_\mu
X^mC_m^{(1)}-\partial_\mu Z \ .
\end{equation}
As the next step we use the fact that
\begin{eqnarray} \det
(\calg_{\mu\nu}+e^{2\phi}Y_\mu Y_\nu)
&=&\det \calg (1+
e^{2\phi}\calgi^{\mu\nu}Y_\mu
Y_\nu ) \nonumber \\
\end{eqnarray}
and introduce an auxiliary variable $v$
so that
\begin{eqnarray}
-\int d^4\xi V\sqrt{-\det\bA}=
\frac{1}{2} \int d^4\xi (\frac{1}{v}
V^2\det \bA-v )= \nonumber \\
=\frac{1}{2}\int d^4\xi
(\frac{e^{-2\phi}}{v} V^2\det \calg+
\frac{V^2}{v} \det\calg Y \calgi Y -v )
\ . \nonumber
\\
\end{eqnarray}
Further we interpret  $Y$ as  one
form on the world-volume of M3-brane so
 that $dZ=C^{(1)}-Y$. Then using the fact
that $d^2Z=0$ we obtain
\begin{eqnarray}\label{dCY}
d(C^{(1)}-Y)=0 \ .  \nonumber \\
\end{eqnarray}
We can consider $Y$ as an independent
field  when we add to the action an
expression
\begin{equation}\label{addA}
-\int d^4\xi F\wedge V(T) dT\wedge
(C^{(1)}-Y) \ ,
\end{equation}
where  $F=dA$ and  where $A$ is
one-form on the world-volume of
M3-brane. Then it is easy to see that
the variation of (\ref{addA}) with
respect to $A$ implies (\ref{dCY})
\begin{eqnarray}
\frac{\delta }{\delta A}[- \int d^4\xi
F\wedge  dT\wedge  (C^{(1)}-Y)]= - \int
d^4\xi
\delta A  dT\wedge d(C^{(1)}-Y)=0 \nonumber \\
\end{eqnarray}
using the fact that
$d(V(T)dT)=V'dT\wedge dT=0$. As the
next step we use (\ref{ansdr}) together
with the definition of $Y$  given in
(\ref{calgY}) to write WZ term as
\begin{eqnarray}
S_{WZ}&=&-\frac{1}{3!}\int d^4\xi
V(T)\epsilon^{\mu_1\dots\mu_4}\partial_{\mu_1}T
\hC_{\mu_2\dots\mu_4}=\nonumber \\
&=&-\frac{1}{3!}\int d^4\xi V(T)
\epsilon^{\mu_1\dots\mu_4}\partial_{\mu_1}T
[ C_{mnk}^{(3)}
\partial_{\mu_2}X^m\partial_{\mu_3}
X^n\partial_{\mu_4}X^k+\nonumber \\
&+&
3B_{mn}\partial_{\mu_2}X^m\partial_{\mu_3}
X^n C^{(1)}_{\mu_4}]+\nonumber \\
&+&\frac{1}{2}\int d^4\xi
\epsilon^{\mu_1\dots\mu_4}
V(T)\partial_{\mu_1}T B_{mn}
\partial_{\mu_2}X^m\partial_{\mu_3}X^n
Y_{\mu_4} \ .  \nonumber \\
\end{eqnarray}
Finally we integrate out $Y$ from the
action and we obtain
\begin{eqnarray}
Y_\nu=
-\frac{v}{\det\calg}\calg_{\nu\mu}
\epsilon^{\mu\mu_1\mu_2\mu_3}
(F_{\mu_1\mu_2}+B_{\mu_1\mu_2})\partial_{\mu_3}T
\nonumber \\
\end{eqnarray}
that we insert back to the action and
we obtain
\begin{eqnarray}\label{Sv}
S&=&\frac{1}{2} \int d^4\xi\left[
\frac{e^{-2\phi}V^2}{v}\det\calg
-\frac{v}{\det\calg}
(\epsilon^{\mu\mu_2\mu_3\mu_4}
(B+F)_{\mu_2\mu_3}\partial_{\mu_4}T)
\calg_{\mu\nu}
(\epsilon^{\nu\nu_2\nu_3\nu_4}
(B+F)_{\nu_2\nu_3}\partial_{\nu_4}T)\right]-
\nonumber \\
&-&\frac{1}{3!}\int d^4\xi
\epsilon^{\mu_1\dots\mu_4}
V\partial_{\mu_1}T C^{(3)}_{\mu_2\dots\mu_4}
-\frac{1}{2}\int d^4\xi
\epsilon^{\mu_1\dots\mu_4}
V\partial_{\mu_1}
T(F+B)_{\mu_2\mu_3}C^{(1)}_{\mu_4} \ .
\nonumber
\\
\end{eqnarray}
We can derive an alternative form of
the action if we integrate $v$ from
(\ref{Sv}) and hence
\begin{eqnarray}
S&=&-\int d^4\xi e^{-\Phi} V
\sqrt{-\epsilon^{\mu\mu_2\mu_3\mu_4}
(B+F)_{\mu_2\mu_3}\partial_{\mu_4}T
\calg_{\mu\nu}
\epsilon^{\nu\nu_2\nu_3\nu_4}
(B+F)_{\nu_2\nu_3}\partial_{\nu_4}T}-
\nonumber \\
&-&\frac{1}{3!}\int d^4\xi
\epsilon^{\mu_1\dots\mu_4}
V\partial_{\mu_1}T C^{(3)}_{\mu_2\dots\mu_4}
-\frac{1}{2}\int d^4\xi
\epsilon^{\mu_1\dots\mu_4}
V\partial_{\mu_1}
T(F+B)_{\mu_2\mu_3}C^{(1)}_{\mu_4} \ .
 \nonumber
\\
\end{eqnarray}
We see that the resulting action is
 different from the
standard unstable D3-brane action and
hence it seems to us that the naive
form of the action (\ref{M3act}) is not
the correct one that should correspond
to M3-brane action. For that reason we
suggest an alternative M3-brane action
with the main property
 that its
dimensional reduction leads to
effective action for an unstable
D3-brane.
\section{Proposal for Non-BPS M3-brane}
\label{fourth}
In order to construct M3-brane action
whose direct dimensional reduction
leads to D3-brane action in type IIA
theory we  proceed in a similar way as
in the case of  the construction of the
massive M-theory KK-monopole action
\cite{Bergshoeff:1998ef}. Following
this paper we consider  the target
space background corresponding to
11-dimensional supergravity with
 the  field content
($\hg_{MN} \ , \hC_{MNK}$) and that
 has an isometry generated by a
Killing vector $\hk^M(X)$ such that the
Lie derivative of all target space
fields with respect to $\hk^M$ vanish
\begin{equation}
\mL_{\hk}\hC=\mL_{\hk}\hg=0 \ .
\end{equation}
Then, with analogy with
\cite{Bergshoeff:1998ef} we propose
 DBI and WZ parts of  M3-brane
effective action in the form
\begin{equation}\label{acgNP}
S_{DBI}=-\int d^4\xi V(T)
|\hk|^{\frac{1}{2}}\sqrt{-\det
\bA}
\end{equation}
and
\begin{eqnarray}\label{WZNP}
S_{WZ}= -\int  d^4\xi
V(T)\epsilon^{\mu_1\dots \mu_4}
\partial_{\mu_1}T\hmK_{\mu_2\mu_3\mu_4}
 \ , \nonumber \\
 \end{eqnarray}
where
\begin{eqnarray}\label{defbA}
\bA_{\mu\nu}&=& \hg_{MN} \hD_\mu \hX^M
\hD_\nu \hX^N+\frac{1}{|\hk|}
\hmF_{\mu\nu}+\frac{1}{|\hk|}\partial_\mu
T\partial_\nu T \ , \nonumber \\
\hk^2&=&\hk^M\hk^N\hg_{MN} \ , \quad
\hk^2=|\hk|^2 \ ,  \nonumber \\
\hmF_{\mu\nu}&=&\partial_\mu \hb_\nu-
\partial_\nu \hb_\mu+
\partial_\mu \hX^M\partial_\nu \hX^N
(i_{\hk}\hC)_{MN} \ , \nonumber \\
 \hD_\mu X^M&=&
\partial_\mu \hX^M-\hA_\mu \hk^M \ , \quad
\hA_\mu=\frac{1}{|\hk|^2}
\partial_\mu  \hX^M \hk_M \ ,
\nonumber \\
  \hmK_{\mu_2\mu_3\mu_4}&=&\partial_{\mu_2}\homega_{\mu_3\mu_4}^{(2)}-\partial_{\mu_3}
\homega_{\mu_2\mu_4}^{(2)}+
\partial_{\mu_4}\homega^{(2)}_{\mu_2\mu_3}+
\nonumber \\
&+&
\frac{1}{3!}\hC_{KMN}\hD_{\mu_2}\hX^K
\hD_{\mu_3}\hX^M
\hD_{\mu_4}\hX^N+\frac{1}{2!}\hA_{\mu_2}(\partial_{\mu_3}
\hb_{\mu_4}-\partial_{\mu_4}\hb_{\mu_3})
\
, \nonumber \\
\end{eqnarray}
and where, following
\cite{Bergshoeff:1998ef}
 we introduced non-propagating
world-volume $2$-form $\homega$ that describes
tension of M3-brane.

Let us now discuss the gauge symmetries
of the actions (\ref{acgNP}) and
(\ref{WZNP}). It is easy to see that
they are  invariant under local
isometry transformation labeled by
$\hsigma^{(0)}$
\begin{eqnarray}
\delta \hX^M &=&-\hsigma^{(0)}\hk^M \ ,
\quad \delta \hg_{MN}
=-\hsigma^{(0)}\partial_K
\hg_{MN}\hk^K \ ,  \nonumber \\
\delta \hk^M
&=&
-\hsigma^{(0)}\partial_K\hk^M\hk^K \ ,
\quad \delta |\hk|^2=0 \ ,
 \nonumber \\
\delta \hk_M &=&\hsigma^{(0)}
\partial_M\hk^K\hk_K  \ , \quad
 \delta
\hA_\mu
= -\partial_\mu \hsigma^{(0)} \ ,
\nonumber \\
\delta \homega_{\mu\nu}&=&
-\frac{1}{2}[\partial_\mu \hsigma^{(0)}
\hb_\nu -\partial_\nu
\hsigma^{(0)}\hb_\mu] \ ,  \nonumber \\
\delta \hC_{KLM}&=&
-\hsigma^{(0)}\hk^N\partial_N
\hC_{KLM} \ .  \nonumber \\
\end{eqnarray}
Further, the action is invariant under
local gauge transformation labeled by
$\hrho^{(0)}$
\begin{eqnarray}
\delta \hb_\mu=
\partial_\mu  \hrho^{(0)} \ , \quad
\delta \hX^M=0 \ , \quad \delta T=0 \ , \quad
\delta \hmF_{\mu\nu}=0 \ . \nonumber \\
\end{eqnarray}
Finally the WZ action is also invariant
under the gauge transformation of   two
form $\homega_{\mu\nu}$
\begin{eqnarray}
\delta \homega_{\mu\nu}=
\partial_\mu\hrho^{(2)}_{\nu}
-\partial_\nu\hrho^{(2)}_\mu \ .
\nonumber
\\
\end{eqnarray}
Now we show that for the
background (\ref{ansdr})
the actions (\ref{acgNP}),(\ref{WZNP})
describe  non-BPS D3-brane action in
type IIA theory.
%
In fact, when we use  (\ref{ansdr}) in
(\ref{defbA}) we obtain
\begin{eqnarray}\label{ansDR}
\hA_\mu &=&
-C^{(1)}_m\partial_\mu
X^m+\partial_\mu Z \ ,  \quad \hD_\mu
X^m=\partial_\mu X^m, \quad  \hD_\mu Z=
|\hk|=e^{\frac{2}{3}\phi} \ ,  \nonumber \\
\bA_{\mu\nu}&=&
e^{-\frac{2}{3}\phi}
[g_{mn}\partial_\mu X^m\partial_\nu X^n+
F_{\mu\nu}+B_{mn}
\partial_\mu X^m\partial_\nu X^n+\partial_\mu T\partial_\nu T]
\ . \nonumber \\
\end{eqnarray}
Then if we insert this result into the
action (\ref{acgNP}) we obtain
\begin{equation}
S=-\int d^4\xi e^{-\phi} V(T)\sqrt{
-\det (g_{mn}\partial_\mu X^m\partial_\nu X^n+
F_{\mu\nu}+B_{mn}
\partial_\mu X^m\partial_\nu X^n+\partial_\mu T\partial_\nu T)}
\end{equation}
that is clearly the correct  form of
the D3-brane   effective action.

As the next step we perform dimensional reduction
in case of WZ term (\ref{WZNP}). In fact, using
(\ref{ansDR}) we find
\begin{eqnarray}
S_{WZ}&=&  \int d^4\xi
V(T)\epsilon^{\mu_1\dots\mu_4}
\partial_{\mu_1}T
[\frac{1}{3!}C_{mnk}^{(3)}
\partial_{\mu_2}X^m\partial_{\mu_3}X^n
\partial_{\mu_4}X^k-\nonumber \\
&-&\frac{1}{2!}
(\partial_{\mu_2}b_{\mu_3}-\partial_{\mu_3}
b_{\mu_2}+B_{mn}\partial_{\mu_2}
X^m\partial_{\mu_3}X^n)C_{k}^{(1)}\partial_{\mu_4}X^k+
\nonumber \\
&+&(-\partial_{\mu_2}\omega_{\mu_3\mu_4}
+\frac{1}{2}\partial_{\mu_2}Z
(\partial_{\mu_3}b_{\mu_4}-
\partial_{\mu_4}b_{\mu_3}))] \
\nonumber \\
\end{eqnarray}
that is again correct form of the
Wess-Zumino term for  non-BPS D3-brane
that now contains additional
contribution from
  non-propagating
two form $\hat{c}^{(2)}_{\mu\nu}=
-\homega^{(2)}_{\mu\nu}+Z \frac{1}{2}
(\partial_\mu B_\nu-\partial_\nu
B_\mu)$.

In summary, we found the form of
unstable M3-brane action that in
natural way-upon direct dimensional
reduction-leads to correct form of
non-BPS D3-brane action. However it
will be interesting to study this
unstable M3-brane action further. In
the next section
we  implement double
dimensional reduction of given action
and try  to identify resulting object.
\section{Double Dimensional Reduction}
\label{fifth}
 In this section we briefly discuss the
  double dimensional reduction of an unstable
M3-brane, following
\cite{Bergshoeff:1998ef}. For
simplicity we restrict ourselves to the
analysis of DBI part of the action only
however its extension to WZ term is
straightforward.

 In order to
perform the double dimensional
reduction we have to introduce an extra
isometry for the background that is
generated by a Killing vector $\hh$ and
then wrap one direction of the
M3-brane, $\xi^3$ around this new
compact direction
\begin{equation}
\partial_\xi \hX^M=\hh^M \ .
\end{equation}
Generally we have two different
isometries: one is given by $\hh$ that
is a direction tangent to a
world-volume and the other one is given
by $\hk$ in direction transverse to the
world-volume. Then the action
(\ref{acgNP}) will be invariant under
both isometries when $\mL_{\hh}\hk=0$.
In other words we can find a coordinate
system adapted coordinates in both
isometries:
\begin{equation}
\hh^M=\delta^{My} \ , \quad \hk^M=
\delta^{Mz} \ .
\end{equation}
In this double dimensional reduction
the Killing vector $\hk$ becomes after
reduction the Killing vector associated
to the isometry of the space transverse
to the world-volume of an unstable
M3-brane
\begin{equation}
\hk^y=0 \ ,  \quad \hk^m=k^m \ .
\end{equation}
We also use the following reduction
rules for the background fields
\begin{eqnarray}
\hg_{mn}&=&e^{-\frac{2}{3}\phi}
g_{mn}+e^{\frac{4}{3}\phi}C_m^{(1)}
C_n^{(1)} \ , \nonumber \\
\hg_{my}&=&-e^{\frac{4}{3}}C^{(1)}_m \
, \quad \hg_{yy}=e^{\frac{4}{3}\phi} \
,
\nonumber \\
\hC_{kmn}&=&C_{kmn}^{(3)} \ , \quad
\hC_{mn
y}=B_{mn} \ , \nonumber \\
(i_{\hk}\hC)_{my}&=& k^n \hC_{myn}=-
\hC_{mny}k^n= -B_{mn}k^n
=-(i_k B)_m \ .  \nonumber \\
\end{eqnarray}
For simplicity we restrict ourselves to
the background with zero $C^{(1)}_m$.
Then we obtain
\begin{eqnarray}
\hk^2
=e^{-\frac{2}{3}\phi} k^2
\ , \quad
k^2=g_{mn}k^m k^n \nonumber \\
\end{eqnarray}
so that
\begin{eqnarray}
\hA_\alpha=
\frac{1}{k^2}\partial_\alpha X^m
g_{mn}k^n \equiv A_\alpha \ , \quad
\hA_3=0 \ .
\end{eqnarray}
Further, we have
\begin{eqnarray}
\hb_\alpha &=& b_\alpha \ , \quad
\hb_3=
\omega^{(0)} \ , \nonumber \\
\hD_\alpha \hX^m &=&
\partial_\alpha X^m-
A_\alpha k^m\equiv D_\alpha X^m \ ,
\quad \hD_3\hX^m=0 \ , \nonumber \\
 \hmF_{\alpha\beta}&\equiv&\mF_{\alpha\beta}=
\partial_\alpha b_\beta-\partial_\beta
b_\alpha+\partial_\alpha
X^m\partial_\beta X^n C_{mnp}^{(3)}k^p
\nonumber \\
\hmF_{\alpha 3}&=&-\hmF_{3\alpha}=
\partial_\alpha \omega^{(0)}
-\partial_\alpha X^m B_{mn}k^n \equiv
\mK_\alpha^{(1)} \ , \nonumber \\
\hD_\alpha Y&=& 0 \ , \quad  \hD_3 Y= 1
\ .
 \nonumber \\
\end{eqnarray}
 Then we have
\begin{eqnarray}
\bA_{\alpha\beta}&=&
e^{-\frac{2}{3}\phi}(
 D_\alpha X^m
D_\beta X^n g_{mn}+\frac{e^{\phi}}
{|k|}\mF_{\alpha\beta}
+\frac{e^{\phi}}{|k|}\partial_\alpha
T\partial_\beta T) \ ,  \nonumber \\
\bA_{33}&=&
e^{\frac{4}{3}\phi} \ , \quad
\bA_{\alpha 3}=-\bA_{3\alpha}=
\frac{e^{\frac{1}{3}\phi}}
{|k|}\mK_\alpha^{(1)}
\nonumber \\
\end{eqnarray}
and hence
\begin{eqnarray}
\det \bA_{\mu\nu}&=&\det
(\bA_{\alpha\beta}-\bA_{\alpha
3}\bAi^{33}\bA_{3\beta})\bA_{33}= \nonumber \\
&=&e^{-\frac{2}{3}\phi} \det( D_\alpha
X^m D_\beta X^m g_{mn}+\frac{e^{\phi}}
{|k|}\mF_{\alpha\beta}
+\frac{e^{\phi}}{|k|}\partial_\alpha
T\partial_\beta
T+\mK^{(1)}_\alpha\mK^{(1)}_\beta)
\nonumber \\
\end{eqnarray}
so that DBI part of the action takes
the form
\begin{eqnarray}\label{NSiia}
S=-  \int d^3\xi V'(T)
e^{-\frac{1}{2}\phi} |k|^{1/2}
\sqrt{-\det ( D_\alpha X^m D_\beta X^m
g_{mn}+\frac{e^{\phi}}
{|k|}\mF_{\alpha\beta}
+\frac{e^{\phi}}{|k|}\partial_\alpha
T\partial_\beta
T+\mK^{(1)}_\alpha\mK^{(1)}_\beta)} \ ,
\nonumber \\
\end{eqnarray}
where
\begin{equation}
V'(T)=\int d\xi^3 V(T) \ .
\end{equation}
It is natural to interpret the double
dimensional reduction of an unstable
M3-brane as an unstable
three-dimensional object in type IIA
theory whose existence can be deduced
as follows. type IIB theory contains
non-BPS D2-brane with tension
$\tau_{D2}=\sqrt{2}\frac{1}{g_s
l_s^3}$. Under S-duality  rules given
as
\begin{equation}
g_s'=\frac{1}{g_s}  \ , \quad
l_s'=g_s^{1/2}l_s
\end{equation}
D2-brane tension transforms as
\begin{equation}
\tau_{D2}\rightarrow \frac{\sqrt{2}}{l'^3_s
g'^{1/2}_s}=\tau_{NS2} \ ,
\end{equation}
where the resulting object can be
interpreted as unstable NS2-brane.

Let us now  perform T-duality between
type IIA and type IIB theories with
T-duality rules for $g'_s, l'_s, R'$
\begin{equation}
\tilde{g}_s=\frac{g'_sl'_s}{R'} \ ,
\quad \tilde{l}_s=l'_s \ , \quad
\tilde{R}=\frac{l'^2_s}{R'} \ ,
\end{equation}
where symbols with tilde correspond to
type IIA theory while symbols with bar
correspond to type IIB theory. Further
we presume that  T-duality circle is transverse to
the world-volume of type IIA NS2-brane so
that its tension transforms as
\begin{equation}
\tau_{NS2}\rightarrow
\frac{1}{\tilde{l}^3_s
\tilde{g}_s^{1/2}
}\frac{\tilde{R}^{1/2}}{\tilde{l}^{1/2}_s} \ .
\end{equation}
However this agrees with the
tension evaluated from the action
(\ref{NSiia}) on condition that
T-duality circle with radius
$\tilde{R}$  coincides with
the isometry direction that
is transverse to 3-brane in type IIA
theory  since
then $|k|^{1/2}=\frac{\tilde{R}^{1/2}}{l^{1/2}_s}$.

In summary, the double dimensional reduction
of the action (\ref{acgNP}) leads to the
three dimensional object that also naturally
arises as a consequence of chains of dualities
in type IIA and type IIB theories.
We mean that this fact again
 supports our proposed form of
M3-brane action (\ref{acgNP}).

\section{Tachyon Kink on
M3-Brane}\label{sixth}
 We found in the
previous section an action for unstable
M3-brane action that reduces to non-BPS
D3-brane action when we perform direct
dimensional reduction. This is very
attractive property of given action.
However we would like also study the
tachyon condensation on given M3-brane
and identify the resulting object.

To do this we closely follow analysis
presented in  section (\ref{second}) so
that we can be brief. We start with the
ansatz
\begin{eqnarray}\label{ans}
 T &=&f(a(\xi^3 - t(\xi^\alpha))) , \quad
 \nonumber \\
\hX^M&=&\hx^M(\xi^\alpha) \ , \quad
\hb_\alpha= \hb_\alpha(\xi^\alpha) \ ,
\quad \hb_3=0 \ ,
\nonumber \\
\end{eqnarray}
where $\xi^\alpha, \alpha,\beta=0,1,2$
are coordinates tangential to the
world-volume of the kink. The function
$f$ introduced in (\ref{ans}) has the
same properties as in (\ref{fu}).

 The first goal of our
analysis is to show that the actions
(\ref{acgNP}) and (\ref{WZNP})
evaluated on the ansatz (\ref{ans})
leads to the following action
\begin{eqnarray}\label{Sm2a}
S&=&S^{M2g}_{DBI}+S^{M2g}_{WZ} \ , \nonumber \\
S^{M2g}&=& -T_{M2} \int d^3\xi
\sqrt{-\det\ba} \ , \quad
S^{WZg}=-T_{M2} \int d^3\xi
\epsilon^{\alpha\beta\gamma}
\hmK_{\alpha\beta\gamma} \ , \nonumber
\\
& & \ba_{\alpha\beta}=\hD_\alpha
\hx^M\hD_\beta \hx^N
\hg_{MN}+\frac{1}{|\hk|}\hat{f}_{\alpha\beta}
 \ ,
 \nonumber \\
\hat{f}_{\alpha\beta}&=&\partial_\alpha
\hb_\beta-\partial_\beta \hb_\alpha+
\partial_\alpha \hx^M\partial_\beta
\hx^N (i_{\hk}\hC)_{MN} \ , \nonumber
\\
 \hmK_{\alpha\beta\gamma}
 &=&\partial_{\alpha}\homega_{\beta\gamma}^{(2)}-\partial_{\beta}
\homega_{\alpha\gamma}^{(2)}+
\partial_{\gamma}\homega^{(2)}_{\alpha\beta}+
\nonumber \\
&+&
\frac{1}{3!}\hC^{(3)}_{KMN}\hD_{\alpha}\hx^K
\hD_{\beta}\hx^M
\hD_{\gamma}\hx^N+\frac{1}{2!}\ha_{\alpha}(\partial_{\beta}
\hb_{\gamma}-\partial_{\gamma}\hb_{\beta})
\ . \nonumber \\
 \end{eqnarray}
 We will study properties of this
 action in the next section. Now
  we determine from (\ref{Sm2a}) the
equations of motion for $\hx^M$
\begin{eqnarray}\label{eqx}
&-&\frac{1}{2} \left(\frac{\delta
\hg_{MN}}{\delta \hx^K}
\partial_\mu \hx^M\partial_\nu \hx^N
-\frac{\delta |\hk|}{\delta \hx^K}
\frac{1}{|\hk|^2}\hat{f}_{\alpha\beta}
+\right. \nonumber \\
& & \left. +
\frac{1}{|\hk|}\partial_\alpha
\hx^M\partial_\beta \hx^N \frac{\delta
(i_{\hk}\hC)_{MN}}{\delta\hx^K} \right)
\bai^{\beta\alpha}\sqrt{-\det\ba}+
\nonumber \\
&+&
\partial_\alpha\left[
\hg_{KM}\partial_\beta \hx^M
\bai^{\beta\alpha}_S
\sqrt{-\det\ba}\right]+\nonumber \\
&+&\partial_\alpha \left[
\frac{1}{|\hk|}(i_{\hk}\hC)_{KM}
\partial_\beta \hx^M
\bai^{\beta\alpha}_S\sqrt{-\det\ba}\right]
+\tilde{J}_K=0  \ ,
\nonumber \\
\end{eqnarray}
where $\tilde{J}_K=\frac{\delta}{\delta
\hx^K}
 S^{M2g}_{WZ}$.
Further, the equations of motion for
$\hb_\alpha$ take the form
\begin{eqnarray}\label{eqba}
\partial_\alpha[
\bai^{\beta\alpha}_A\sqrt{-\det\ba}]+J_\alpha
=0 \ , \quad
\tilde{J}^\alpha=\frac{\delta }{\delta
\hb_\alpha}S^{M2g}_{WZ}=
\partial_\beta[\epsilon^{\alpha\gamma\beta}
\ha_\gamma] \ .
 \nonumber \\
\end{eqnarray}
As the first step we insert the ansatz
(\ref{ans}) to the matrix $\bA$ and we
get
\begin{eqnarray}\label{bA}
\bA_{\alpha\beta}&=& \hD_{\alpha}\hx^M
\hD_{\beta}\hx^N\hg_{MN}+
\frac{1}{|\hk|}
\hat{f}_{\alpha\beta}+\frac{1}{|\hk|}
a^2f'^2\partial_\alpha t\partial_\beta
t \ , \nonumber \\
 \bA_{\alpha
3}&=&-\frac{1}{|\hk|}a^2f'^2\partial_\alpha
t \ , \bA_{3\beta}=-\frac{1}{|\hk|}
a^2f'^2\partial_\beta t  \ ,
\bA_{33}=\frac{1}{|\hk|} a^2f'^2
\nonumber \\
\end{eqnarray}
Then it is easy to see that
\begin{equation}\label{detBA}
\det \bA_{\mu\nu}= \frac{a^2f'^2}{|\hk|} \det
\ba_{\alpha\beta}
\end{equation}
and also
\begin{eqnarray}
\bAi^{33}&=&\frac{|\hk|}
{a^2f'^2}+\partial_\alpha
t\bai^{\alpha\beta}
\partial_\beta t \ , \quad
\bAi^{\alpha 3}= \bai^{\alpha\gamma}
\partial_\gamma t \ , \nonumber \\
\bAi^{3\beta}&=&
\partial_\gamma t \bai^{\gamma\beta}
 \ , \quad
\bAi^{\alpha\beta}=\bai^{\alpha\beta} \
.
\nonumber \\
\end{eqnarray}
Then if we insert (\ref{detBA}) into
(\ref{acgNP}) and take the limit
$a\rightarrow \infty$ we obtain
\begin{eqnarray}
S_{M3}=-\tau_{M3} \int d\xi^3 d^3\xi
 a
f' V \sqrt{-\det \ba} =-T_{M2} \int
d^3\xi \sqrt{-\det\ba} \ ,
\end{eqnarray}
where
\begin{equation}
T_{M2}=\tau_{M3}\int dx
af'(ax)V(f(ax))= \tau_{M3}\int du V(u)
\ .
\end{equation}
In the same way we can show that when
we insert the ansatz (\ref{ans}) into
(\ref{WZNP}) and take the limit
$a\rightarrow \infty$ we derive the WZ
part of the action given in
(\ref{Sm2a}).

As the next step we show that the
equations of motion for $T$ and $\hX$
are obeyed for the ansatz (\ref{ans})
on condition that the modes $\hx^M$
obey the equations of motion
(\ref{eqx}). In fact, it is easy to
determine from (\ref{acgNP}) the
equation of motion for $T$
\begin{equation}\label{eqTg}
-|\hk|^{\frac{1}{2}}V'(T)\sqrt{-\det\bA}+
\frac{1}{2}
\partial_\mu
\left[|\hk|^{\frac{1}{2}}V(T)\frac{1}{|\hk|}\partial_\nu
T
\bAi^{\nu\mu}_S\sqrt{-\det\bA}\right]+J_T=0
\ ,
\end{equation}
where $J_T=\frac{\delta}{\delta T}
S_{WZ}$ is the source current derived
from varying the WZ term (\ref{WZNP}).
Then, following the same analysis as in
section (\ref{second}) it can be
easily shown that the tachyon
equation of motion are satisfied for
any $t$.

Let us now turn to the equations of
motion for the scalar modes $\hX^M$
that follow from the variation of
(\ref{acgNP}) and (\ref{WZNP})
\begin{eqnarray}\label{eqXg}
&-&\frac{\delta |\hk|^{\frac{1}{2}}}
{\delta \hX^K}V\sqrt{-\det\bA}-
\frac{|\hk|^{\frac{1}{2}}}{2}
V\left(\frac{\delta \hg_{MN}}{\delta
\hX^K}
\partial_\mu \hX^M\partial_\nu \hX^N
-\frac{\delta |\hk|}{\delta \hX^M}
\frac{1}{|\hk|^2}
(\mF_{\alpha\beta}+\partial_\alpha T\partial_\beta T)
+\right. \nonumber \\
& &\left.+ \frac{1}{|\hk|}\partial_\mu
\hX^M\partial_\nu \hX^N \frac{\delta
(i_{\hk}\hC)_{MN}}{\delta\hX^K} \right)
\bAi^{\nu\mu}\sqrt{-\det\bA}+
\nonumber \\
&+&
\partial_\mu\left[|\hk|^{\frac{1}{2}}V
\hg_{KM}\partial_\nu \hX^M
\bAi^{\nu\mu}_S
\sqrt{-\det\bA}\right]+\nonumber \\
&+&
\partial_\mu
\left[|\hk|^{\frac{1}{2}}V
\frac{1}{|\hk|}(i_{\hk}\hC)_{KM}
\partial_\nu \hX^M
\bAi^{\nu\mu}_A\sqrt{-\det\bA}\right]
+J_K=0  \ ,
\nonumber \\
\end{eqnarray}
where $J_K=\frac{\delta S_{WZ}}{\delta
\hX^K}$. Inserting the ansatz
(\ref{ans}) to the equation
(\ref{eqXg}) we obtain
\begin{eqnarray}\label{eqXgi}
&-&Vaf'\frac{1}{2}\left(\frac{\delta
\hg_{MN}}{\delta \hx^K}
\partial_\alpha \hx^M\partial_\beta
\hx^N
 -\frac{\delta
|\hk|}{\delta \hx^M} \frac{1}{|\hk|^2}
\mF_{\alpha\beta}+\right.\nonumber \\
& &
\left.+\frac{1}{|\hk|}\partial_\alpha
\hx^M\partial_\beta \hx^N \frac{\delta
(i_{\hk}\hC)_{MN}}{\delta\hx^K}\right)
\bai^{\beta\alpha}\sqrt{-\det\ba}+
\nonumber \\
&+&Vaf'
\partial_\alpha \left[
\hg_{KM}\partial_\beta \hx^M
(\bai^{\beta\alpha}_S
\sqrt{-\det\ba}\right] +\nonumber \\
&+&aV'\partial_\alpha \left[
\frac{1}{|\hk|}(i_{\hk}\hC)_{KM}
\partial_\nu \hx^M
\bai^{\beta\alpha}_A\sqrt{-\det\ba}\right]+J_K=
\nonumber \\
&=& af'V[eq. \ of \  m.(\ref{eqx})]=0 \  \nonumber \\
\end{eqnarray}
using the fact that
\begin{eqnarray}
J_K=-V(T)\epsilon^{\mu_1\dots\mu_4}
\partial_{\mu_1}T\frac{\delta
\hmK_{\mu_2\dots\mu_4}}{\delta\hx^K}=
-V(f)af'\epsilon^{\alpha\beta\gamma}
\frac{\delta
\hmK_{\alpha\beta\gamma}}{\delta\hx^K}
=af'V\tilde{J}_K \ .  \nonumber \\
\end{eqnarray}
In other words the  dynamics of the
tachyon kink is governed by the
equations of motion (\ref{eqx}).

As the final step we determine from
(\ref{acgNP}) and (\ref{WZNP}) the
equation of motion  for $\hb_\mu$
\begin{equation}\label{eqA}
\partial_\nu
\left[|\hk|^{\frac{1}{2}} V
\frac{1}{|\hk|}\bAi^{\mu \nu}_A
\sqrt{-\det\bA}\right]+J^\mu=0 \ ,
\end{equation}
where $J_\mu=\frac{\delta}{\delta
\hb_{\mu}}
 S_{WZ}$. After some manipulations
we again find that they are obeyed by
the  ansatz (\ref{ans}) on condition
that $\hb_\alpha $ obeys (\ref{eqba}).
 In summary we have shown that the
dynamics of the tachyon kink on the
world-volume of an unstable M3-brane is
governed by the equations of motion
that follow from the variation of the
action (\ref{Sm2a}). Since the form of
this action is not well known let us
now study its properties.
\section{Gauged
M2-Brane Action}\label{seventh}
In this section we review the main
properties of gauged M2-brane
action (\ref{Sm2a}). Firstly we
show that its dimensional reduction
leads to D2-brane action. Further, we
demonstrate   that
 the action (\ref{M2act}) is equivalent to standard
M2-brane action in  flat
background.

We start with the question of
dimensional reduction of the action
(\ref{M2act}). We again presume that
$D=11$ background has a $U(1)$ isometry
with Killing vector field $\hk$
 that in adapted coordinates
$\hx^M = (x^\mu, z)$ for which $\hk
=\frac{\partial}{\partial z} \ ,
\hk^M=\delta^M_Z$ so that the
background takes the form as in
(\ref{ansdr})
so that we obtain
\begin{eqnarray}
\ha_\alpha
&=&-C^{(1)}_m\partial_\alpha
\hx^m+\partial_\alpha z \ , \quad
|\hk|=e^{\frac{2}{3}\phi} \ , \nonumber \\
\hD_\alpha \hx^m &=&\partial_\alpha
\hx^m \ , \quad  \hD_\alpha z=
C_m^{(1)}\partial_\alpha \hx^m=
C_\alpha^{(1)} \ , \quad \nonumber \\
\end{eqnarray}
and hence the matrix $\ba$ defined in
 (\ref{Sm2a}) takes the form
\begin{eqnarray}
\ba_{\alpha\beta}&=&
e^{-\frac{2}{3}\phi}
[g_{mn}\partial_\alpha
x^m\partial_\beta x^n+
f_{\alpha\beta}+B_{mn}
\partial_\alpha x^m\partial_\beta x^n]
\ .
\nonumber \\
\end{eqnarray}
Inserting this result into the action
(\ref{Sm2a}) we obtain
\begin{equation}\label{SDBID2}
S_{DBI}^{M2g}=-T_{M2}\int d^3\xi e^{-\phi}
\sqrt{-\det (g_{\alpha\beta}+
b_{\alpha\beta}+f_{\alpha\beta})}
\end{equation}
that is the correct form of DBI action
for D2-brane. 
In the same way we can show that the WZ
term given in (\ref{Sm2a}) takes the
form
\begin{eqnarray}\label{SWZD2}
S^{M2g}_{WZ}&=& T_3 \int d^3\xi
\epsilon^{\alpha\beta\gamma}
\left(\frac{1}{3!}C_{mnk}^{(3)}
\partial_{\alpha}x^m\partial_{\beta}x^n
\partial_{\gamma}x^k-\right.\nonumber \\
& & \left.-\frac{1}{2!}
(\partial_{\alpha}b_{\beta}-\partial_{\beta}
b_{\alpha}+B_{mn}\partial_{\alpha}
x^m\partial_{\beta}x^n)C_{k}^{(1)}\partial_{\gamma}x^k+
\right.\nonumber \\
& & \left.
+(-\partial_{\alpha}\omega_{\beta\gamma}
+\frac{1}{2}\partial_{\alpha}z
(\partial_{\beta}b_{\gamma}-
\partial_{\gamma}b_{\beta})\right) \nonumber
\\
\end{eqnarray}
This can be considered as the
Wess-Zumino term for D2-brane when we
introduce non-propagating world-volume
two-form $c^{(2)}$ as
\begin{equation}
c^{(2)}_{\alpha\beta}=
-\homega_{\alpha\beta}^{(2)}+
\frac{z}{2}[\partial_\alpha b_\beta-
\partial_\beta b_\alpha] \ .
\end{equation}

Finally we show that the action
(\ref{M2act}) is equivalent to standard
M2-brane action
in the flat eleven dimensional
space-time with $\hg_{MN}=\eta_{MN}\ ,
 \hC_{MNK}=0$. Using $SO(10)$
rotation symmetry of spatial part of
target space-time we can take $\hk^M=
\delta^M_{10}$ so that
$\hA_\alpha=\partial_\alpha \hX^{10}$
so that
\begin{equation}
\hD_\alpha \hX^m=\partial_\alpha\hX^m \ ,
m=0,\dots,9 \ , \quad
\hD_\alpha\hX^{10}=0 \ .
\end{equation}
Then DBI part of M2-brane action takes
the form
\begin{eqnarray}
S_{DBI}^{M2g}&=&-T_{M2}
\int d^3\xi \sqrt{-\det
(\partial_\alpha \hx^m\partial_\beta
\hx^n\eta_{mn}+f_{\alpha\beta})}=
\nonumber \\
&=&-T_{M2}\int d^3\xi \sqrt{-\det
(\partial_\alpha \hx^m \partial_\beta \hx^n
\eta_{mn}+\frac{1}{T_{M2}^2}B_\alpha
B_\beta)}+\frac{1}{2} \int d^3\xi
\epsilon^{\alpha\beta\gamma}B_\alpha f_{\beta\gamma}
\ .
\nonumber \\
\end{eqnarray}
Then if we integrate out $b_\alpha$
from the equation above we obtain
\begin{equation}
\partial_\alpha \epsilon^{\alpha\beta}B_\beta=0
\end{equation}
that can be solved as $B_\alpha=
T_{M2}\partial_\alpha \hx^{10}$. Inserting
this result to the action above and
performing integration by parts we
finally get
\begin{equation}
S=-T_{M2}\int d^3\xi \sqrt{-\det
\partial_\alpha \hx^M\partial_\beta \hx^N} \
\end{equation}
that is standard M2-brane action in
flat background.
\section{Conclusion}\label{cons}
Now  we summarize our results. The
initial goal was to find the action for
 unstable M3-brane that is
related by direct dimensional reduction
to an   unstable D3-brane action and
the singular tachyon kink on its
world-volume leads to stable M2-brane
action. We argued that it is not
possible to obey these two conditions
simultaneously. Then  we suggested
alternative form of M3-brane effective
action that however  exists on
condition that the supergravity
background has Killing isometry that is
gauged on the world-volume of M3-brane.
We also shown that the tachyon kink
solution on its world-volume leads to
M2-brane action that is equivalent to
the standard M2-brane action. This fact
suggests that the proposed M3-brane
action could be starting point for its
non-abelian generalization. It would be
also really interesting to perform  the
same procedure as in paper
\cite{Kluson:2008nw} in this M3-brane
action
 and try to find the relation
with recent works
\cite{Garousi:2008yv,Iengo:2008cq}.
\\
\vskip .2in \noindent {\bf
Acknowledgement:} This work
was supported  by the Czech
Ministry of Education under Contract
No. MSM 0021622409.

\end{document}